\documentclass[fleqn,usenatbib]{mnras}

\usepackage{newtxtext,newtxmath}

\usepackage[T1]{fontenc}

\usepackage{graphicx}	
\usepackage{amsmath}	

\newcommand{\red}[1]{\textcolor{black}{#1}}

\newcommand{\be}{\begin{equation}}
\newcommand{\ee}{\end{equation}}
\newcommand{\beq}{\begin{equation}}
\newcommand{\eeq}{\end{equation}}
\newcommand{\bea}{\begin{eqnarray}}
\newcommand{\eea}{\end{eqnarray}}


\newcommand{\bx}{{\bf{x}}}
\newcommand{\by}{{\bf{y}}}

\newcommand{\nn}{{\nonumber}}

\usepackage[stable]{footmisc}

\newcommand{\al}{\alpha}

\newcommand{\ka}{\kappa}

\renewcommand{\th}{\theta}

\newcommand{\si}{\sigma}
\newcommand{\Si}{\Sigma}

\newcommand{\ra}{\rightarrow}

\newcommand\ees{\end{eqnarray}}
\newcommand\bees{\begin{eqnarray}}

\title[Lensed TDEs]{Lensing of gravitational waves from tidal disruption events}
\author[M.Toscani et al.]{
Martina Toscani$^{1}$\thanks{E-mail: martina.toscani@l2it.in2p3.fr},
Elena M. Rossi,$^{2}$
Nicola Tamanini,$^{1}$
and Giulia Cusin$^{3, 4}$
\\
$^{1}$ Laboratoire des 2 Infinis - Toulouse (L2IT-IN2P3), Université de Toulouse, CNRS, UPS, F-31062 Toulouse Cedex 9, France\\
$^{2}$ Leiden Observatory, Leiden University, PO Box 9513, 2300 RA, Leiden, the Netherlands\\
$^{3}$ Institut d’Astrophysique de Paris, Sorbonne Université, CNRS, UMR 7095, 98 bis bd Arago, 75014 Paris, France\\
$^{4}$ Universit\'e de Gen\'eve, D\'epartement de Physique Th\'eorique and Centre for Astroparticle Physics, 24 quai Ernest-Ansermet, CH-1211 Gen\'eve 4, Switzerland 
}

\date{Accepted XXX. Received YYY; in original form ZZZ}

\pubyear{2022}

\begin{document}
\label{firstpage}
\pagerange{\pageref{firstpage}--\pageref{lastpage}}
\maketitle

\begin{abstract}
In this work, we investigate
the effect of gravitational lensing on the gravitational wave (GW) signals of a population of tidal disruption events (TDEs).
We estimate the number of lensed-magnified signals that we expect to detect with future space-based GW observatories, in particular
LISA and DECIGO.
We model the lens distribution using an hybrid approach that combines semi-analytic methods with numerical results from ray tracing simulations.
We divide the TDE population in two classes, \textit{nuclear TDEs} (main sequence stars tidally disrupted by massive black holes in the cores of galaxies) and \textit{globular TDEs} (white dwarfs tidally disrupted by intermediate mass black holes in globular clusters). 
We find that, even considering the effect of lensing, LISA will not be able to observe any TDEs, while DECIGO could detect $\sim$16 strongly lensed ($\mu>2$) globular TDEs and $\sim$135 strongly lensed nuclear TDEs, over an observational period of 10 years.
Our results reveal the role that lensing will play in future deci-Hertz GW observatories, indicating exciting multi-messenger opportunities with TDEs but at same time signalling the need to develop adequate data analysis techniques to correctly reconstruct the astrophysical properties of the source.

\end{abstract}




\section{Introduction}
Stars orbiting around a massive black hole (BH) can be shred into pieces due to tides induced by the BH's gravitational field.
We call these extreme phenomena tidal disruption events (TDEs; \citealt{Luminet:86aa,Rees:88aa,Phinney:89aa}; for a recent review see \citealt{Rossi:20aa}). Thanks to the bright  electromagnetic (EM) flares produced by the stellar debris during the following accretion, TDEs have been one of the most powerful ways to reveal the presence of otherwise quiescent massive BHs. To date, we have around 100 observations of these events, in different bands of the EM spectrum (see for recent reviews: \citealt{Saxton:20} for X-ray; \citealt{vanvelzen:2020} for optical; \citealt{Alexander:20} for radio and all the references therein).\\
\indent Besides being multiband emitters in the EM spectrum, TDEs are potential multimessenger sources. Recently, a few TDEs have been claimed to be associated to observed astrophysical neutrinos \citep{Stein:22aa,Reusch:22aa} as theoretically expected in the presence of jets \citep{Hayasaki:21aa}. Moreover, TDEs are also predicted to emit gravitational waves (GWs).
In particular, we can distinguish between three main GW contributions: GWs due to the internal stellar mass quadrupole, generated by the stretching and compressing action of the BH tidal forces \citep[see, e.g,][]{Stone:13aa}; GWs associated to the BH-star system mass quadrupole, emitted during the disruption phase \citep[see, e.g.,][]{Kobayashi_2004, Toscani:20aa, Toscani:22aa}; GWs produced at later stages, along the circularization process \citep{Toscani:22aa} and in presence of an accretion disc \citep[see, e.g.,][]{Toscani:19aa}. For standard values of the parameters involved, the strongest gravitational contribution is the burst emitted during the disruption phase, that has typical frequencies in the range $10^{-3}-10^{-2}\,\text{Hz}$.\\
\indent Being low frequency GW sources, TDEs could be detected by future space-based interferometers such as the Laser Interferometer Space Antenna \citep[LISA - ][]{Amaro-Seoane:17aa, Amaro:22aa}, currently scheduled for launch in the mid-2030s, and the proposed deci-Hertz Gravitational Observatory \citep[DECIGO - ][]{Sato:17aa}.
A detailed study about TDEs detectability with these next-generation detectors has been carried out in \citealt{Pfister:21aa}.
Their work shows that while detection of individual TDEs by LISA seems unlikely, these events are promising sources for deci-Hz observatories. Future instruments with a DECIGO-like sensitivity (e.g., DECIGO, BBO - \citealt{Harry:06aa}, ALIA - \citealt{baker:19aa}, DO - \citealt{Sedda:2019uro,Sedda:2021yhn}) could observe hundreds of thousands TDEs per year.\\
\indent Given these expectations, it becomes relevant to assess the effects that a distribution of lenses produces on the GW emission from a TDE population. We refer to gravitational lensing \citep[see, e.g.,][]{Schneider:92aa} when a massive object (i.e. the \textit{lens}), which lies along the line of sight between the observer and the source, curves the surrounding space-time, causing the signal to deviate from its original path. This effect has interesting consequences: for example it may (de-)magnify the signal or produce multiple images of the source. Moreover, different images typically arrive at the detector at different times (time-delay effect) and they interfere if the duration of the signal is larger than the typical delay in the time of arrival. \\
\indent In this work, we study for the first time the effect of gravitational lensing on a TDE population and provide estimates on the expected number of observed lensed-magnified TDEs. We consider both LISA, for which a magnified TDE could be the only way to have a signal above the detectability threshold, but also DECIGO, for which the ability to distinguish lensed TDEs would avoid errors in the reconstruction of the parameters describing the source (e.g., distance and mass), as well as provide additional information on the astrophysical properties of the source and lens populations.
We perform this investigation dividing TDEs in two different classes: \textit{nuclear TDEs}, where we consider main sequence (MS) stars disrupted by massive BHs in the cores of galaxies, and \textit{globular TDEs}, where we consider white dwarfs (WDs) tidally disrupted by intermediate mass BHs (IMBHs) located in globular cluster (GCs). The structure of the paper is the following: in Section \ref{sec:2} we illustrate the basis of gravitational lensing, in Section \ref{sec:3} we describe the distribution of lenses and the TDE populations in details, in Section \ref{sec:4} we show and discuss the results and finally in Section \ref{sec:5} we draw our conclusions.\\
\indent Throughout this work, we adopt a $\Lambda$CDM cosmological model, with matter density parameter $\Omega_{\rm m}=0.274$, dark energy density parameter $\Omega_{\Lambda}=0.726$ and Hubble constant $H_0=70\,\text{Km/sMpc}$.

\section{Gravitational lensing in a nutshell}
\label{sec:2}
We want to determine the number of TDEs that, having a given strain amplitude, or rather a given signal to noise ratio (SNR) $\rho$ for a specified interferometer, are significantly magnified by lensing.
\indent Following \citet{Cusin:21aa}, we present semi-analytic formulae which can be applied to an arbitrary lens and source distribution, keeping full control of modeling and transparency of all physical effects. We mainly focus our analysis on strong lensing, working out the distribution of magnification for $\mu>1$, considering LISA and DECIGO. It is indeed true that the lensing induced by the Cosmological Large Scale Structure can also lead to de-magnification ($\mu<1$) of a signal.
However de-magnification from the Large Scale Structure usually does not reach values $\mu\ll 1$, meaning that its contribution should not significantly affect the detection rates of the observed population.
Furthermore the TDE population is well below the detection threshold of LISA, hence de-magnified events will anyway remain undetected.
For these reasons in what follows we ignore the effect of de-magnification.

Our description of strong lensing relies on the geometric optics approximation (we do not describe wave effects such as diffraction and interference). This is a well-justified approximation when looking at TDEs lensed by a population of galaxies. Indeed, given a lens of mass $M_{\rm l}$, diffraction effects are relevant when \citep[see, e.g,][]{Takahashi:2003ix} 
\begin{align}
    M_{\rm l}\lesssim 10^{8}\text{M}_{\odot}\left(\frac{f}{\rm{mHz}}\right)^{-1},
    \label{eq:diffr}
\end{align}
where $f$ is the frequency of the lensed signal. Since GWs by TDEs have typical frequencies in the range $10^{-3}-10^{-2}$ Hz, wave effects can be neglected in our work, as we consider galaxy stellar masses between $10^{9}
-10^{12}\text{M}_{\odot.}$.

To predict the number of magnified TDEs observable with a given instrument, we need to take the following steps: 
\begin{enumerate}
    \item choose a lens model, and a model to describe the lens distribution and the population of sources; 
    \item derive the probability density function (PDF) for a generic source at redshift $z_{\rm s}$ to be amplified more than $\mu$ by the population of foreground lenses;
    \item convolve the magnification PDF with the population of observable events for a given instrument,  in presence of magnification $\mu$. 
\end{enumerate}

In the following, we model the lenses as singular isothermal spheres (SIS), that we describe in detail in Section \ref{sec:sis}. While SIS are not very realistic when considering lensing by individual clusters, they are sufficient for statistical purposes. In particular, the main advantage of the SIS model is that it can be studied analytically, which allows us to have a better (even tough idealised) comprehension of the physics behind lensing. To follow a more realistic approach in addition to strong lensing, described by our semi-analytic approach, we also consider the contribution from weak lensing due to the gravitational potential of the Large Scale Structure.  
To derive numerical results, we use ray-tracing simulations by \citet{Takahashi:11aa}, which include both weak lensing contribution, and strong lensing tails in the magnification PDF.

\subsection{SIS and gravitational lensing statistics}

\label{sec:sis}
\begin{figure}
    \centering
    \includegraphics[width=0.38\textwidth] {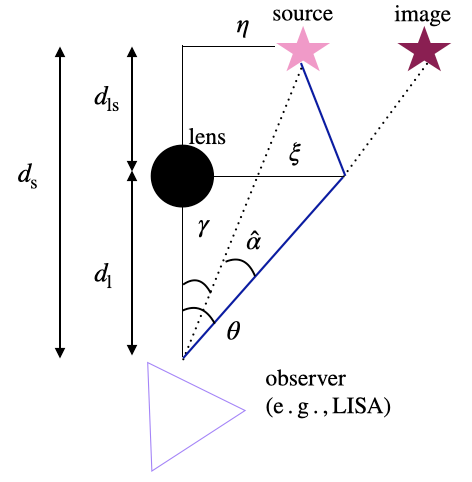}
    \caption{Sketch of the geometry for a SIS lens.}
    \label{fig:geometry}
\end{figure}
In the SIS model all the mass components of the galaxy behave like \textit{particles} of an ideal gas, confined by their combined, spherically symmetric gravitational potential, in thermal equilibrium \citep[see, e.g.,][]{Narayan:96aa}. The mass density of a SIS is described by
\begin{align}
   \varrho (r)=\frac{\sigma^{2}_{\rm v}}{2\pi G r^2},
   \label{eq:mass_density}
\end{align}
where $\sigma_{\rm v}$ is the velocity dispersion in the galaxy, $G$ is the gravitational constant and $r$ is the distance from the center. Integrating Equation \eqref{eq:mass_density} along the line of sight, we get the surface mass density
\begin{align}
    \Sigma (\xi)=\frac{\sigma_{\rm v}^2}{2G\xi},
\end{align}
where $\xi$ is the bi-dimensional vector in the lens plane, called \textit{lens impact parameter}. The geometry of such system is illustrated in Figure \ref{fig:geometry}, where the following elements are shown:
\begin{enumerate}
    \item the source, its lensed image, the observer (e.g., LISA) and the lens,
    \item $\gamma$, angle between the line of sight and the unlensed source,
    \item $\theta$, angle between the line of sight and the lensed image,
    \item $\hat{\alpha}$ deflection angle induced by the presence of the lens,
    \item $d_x$, angular diameter distances between the lens and the source (x=ls), the lens and the observer (x=l), the source and the observer (x=s),
     \item ${\xi}={\theta}d_{\rm l}$, lens impact parameter, in the lens plane,
    \item $\eta={\gamma}d_{\rm s}$, source impact parameter, in the source plane.
\end{enumerate}
We recall \citep[see, e.g,][]{Hogg:99aa} that the angular diameter distance of an object is related to its luminosity distance, $D$, by
\begin{align}
    d= \frac{D}{(1+z)^2},
\end{align}
and the angular diameter distance between two objects at redshifts $z_1$ and $z_2$ is
\begin{align}
    d_{12}=\frac{1}{1+z_2}\left[\chi(z_2)-\chi(z_1)\right]=\frac{1}{1+z_2}\chi(z_{1},z_{2}),
    \label{eq:ang_diff}
\end{align}
where $\chi$ is the comoving distance
\begin{align}
    \chi(z)=\frac{c}{H_0}\int_{0}^{z}\frac{dz'}{E(z')}.
\end{align}
In the above Equation, we have introduced $E(z')=({\Omega_{\rm m}(1+z')^3+\Omega_{\lambda}})^{1/2}$, where $\Omega_{\rm m}$ and $\Omega_{\lambda}$ are the present values of the matter and cosmological constant density contrasts and $c$ is the speed of light in vacuum.\\
\indent The basic quantity for statistical analysis is the cross section of the lens for producing the desired lensing effect (e.g. magnification larger than $\mu$). The corresponding optical depth is the fraction of the sky where, given the lenses, one can place a source and observe this magnification \citep[see, e.g,][]{Kochanek:06aa}.
In the case of the SIS, the area on the source plane in which a source at redshift $z_{\rm s}$ will be magnified $\geq\mu$ is given by $\varsigma_{\rm sis}(\mu,z_{\rm s}, z_{\rm l}, \sigma_{\rm v})=\pi \eta^2(\mu,z_{\rm s}, z_{\rm l}, \sigma_{\rm v})$. The corresponding optical depth  is
\begin{align}
\tau(\mu,z_{\rm s}) =&\int_0^{z_{\rm s}}dz_{\rm l}\frac{dr}{dz_{\rm l}}\textcolor{black}{\left(\frac{d_{\rm l}}{d_s}\right)^2}\int d\sigma_{\rm v} n(\si_{\rm v},z_{\rm l})\varsigma_{\rm sis}(\mu,z_{\rm l},z_{\rm s},\si_{\rm v})\,,
\label{eq:optical_depth}
\end{align}
where $dr$ is the physical length element at redshift $z_{\rm l}$, while $n(\si_{\rm v},z_{\rm l})$ is the physical number density of lenses per bin of $\sigma_{\rm v}$ \citep[][]{Kochanek:06aa}. \\
\indent In the SIS model, there are two lensed images when the source satisfy the following criterion \citep{Schneider:92aa}
\begin{align}
    \gamma<\alpha_{0}=4\pi\frac{\sigma^2_{\rm v}}{c^2}\frac{d_{\rm ls}}{d_{\rm s}},
\end{align}
where $\alpha_{0}$ is usually called \textit{Einstein angle}. In our study we consider this scenario, but we restrict our attention to the primary, i.e.~more magnified, image\footnote{In lensing analyses it is common jargon to refer to the observed signals as ``images'', even though they are not EM signals. In this paper we follow this convention, calling images the lensed GW signals.}, for which we provide the cross section and the explicit final formula for $\tau$ in Appendix  \ref{sec:strong}.
This choice is justified since we expect to see a short burst of GWs from a TDE which comes only from one image.
The second image is in fact delayed in time, with typical time delay of the order of a few months (see e.g.~\citealt{Oguri:2018muv}), much longer than the GW burst itself.
The problem of correctly identifying secondary images, and associating them to their primaries, requires a dedicated data analysis investigation which goes beyond the scope of our study.\\
\indent The probability that a source at redshift $z_{\rm s}$ is magnified more than $\mu$ is
\begin{align}\label{e:P>mu}
P(>\mu,z_{\rm s})=1-\exp({-\tau(\mu,z_{\rm s})})= \int_{\mu}^{+\infty}p(\mu, z_{\rm s})d\mu,
\end{align}
where 
\begin{align}
    p(\mu,z_{\rm s})=-\frac{d\tau}{d\mu} \exp({-\tau(\mu,z_{\rm s})})
\label{eq:prob_dens}
\end{align}
is the magnification PDF.\\
\indent To understand Equation~(\ref{e:P>mu}) note that $d\tau/dz$ can be interpreted as a sort of GW scattering rate leading to magnification larger than $\mu$ (per bin of redshift). Hence the probability for magnification larger than $\mu$ satisfies the differential equation $dP(>\mu,z)/dz = (1-P)d\tau/dz$ with solution (\ref{e:P>mu}). The factor $(1-P)$ is essential to keep the probability normalized also when $\tau$ becomes large.  In the limit of small optical depth, $P(>\mu,z_s)\approx\tau(\mu,z_s)$.\\
\indent As mentioned above, note that our approach does not describe de-magnification which happens when a signal crosses a cosmic under-density. 

\subsection{Gravitational lensing applied to a source population}

We consider a population of sources, that we describe as a function of source redshift $z_{\rm s}$ and SNR $\rho$ and we denote the number of sources per bin of redshift and SNR as $d\mathcal{N}/(d\rho dz_{\rm s})$.  If the magnification is $\mu$, the number of observable sources per bin of $z_{\rm s}$, for a given interferometer, reads
\begin{align}
\frac{d\mathcal{N}(\mu,z_{\rm s})}{dz_{\rm s}}= \int^{\infty}_{\rho_{\rm lim}/\sqrt{\mu}}\frac{d\mathcal{N}}{d\rho dz_{\rm s}}d\rho\,,
\label{eq:dn_dz}
\end{align}
where $\rho_{\rm lim}$ is the threshold above which we have a GW detection. In the rest of the paper, we take $\rho_{\rm lim}=8$.\\
\indent Convolving this quantity with the magnification PDF (Equation \ref{eq:prob_dens}), we get the  total number of observed objects in presence of magnification  
\begin{align}
{\mathcal{N}^{\rm obs}}&=\int_{0}^{z_{\text{max}}}dz_{\rm s}\frac{d\mathcal{N}}{dz_{\rm s}}(z_{\rm s})=\nonumber \\
&\int_{0}^{z_{\text{max}}}dz_{\rm s}\int_{\mu_{\text{min}}}^{+\infty}d\mu p(\mu,z_{\rm s})\frac{d\mathcal{N}(\mu,z_{\rm s})}{dz_{\rm s}}\,,
  \label{eq:dn_obs}
\end{align}
where $z_{\text{max}}$ corresponds to the maximum redshift at which we expect to find sources and $\mu_{\text{min}}$ is the minimum value of magnification considered. The probability that if an instrument sees a source from redshift $z_{\rm s}$ this is magnified more than $\mu$ is given by 
\begin{align}
    \mathcal{P}(z_{\rm s},\mu)=\mathcal{C}\int_{\mu}^{\infty}p(\mu',z_{\rm s})\frac{d\mathcal{N}(\mu',z_{\rm s})}{dz_{\rm s}}d\mu'\,,
    \label{eq:prob_mag}
\end{align}
where $\mathcal{C}$ is a normalization constant.

\begin{figure}
    \centering
    \includegraphics[width=0.5\textwidth] {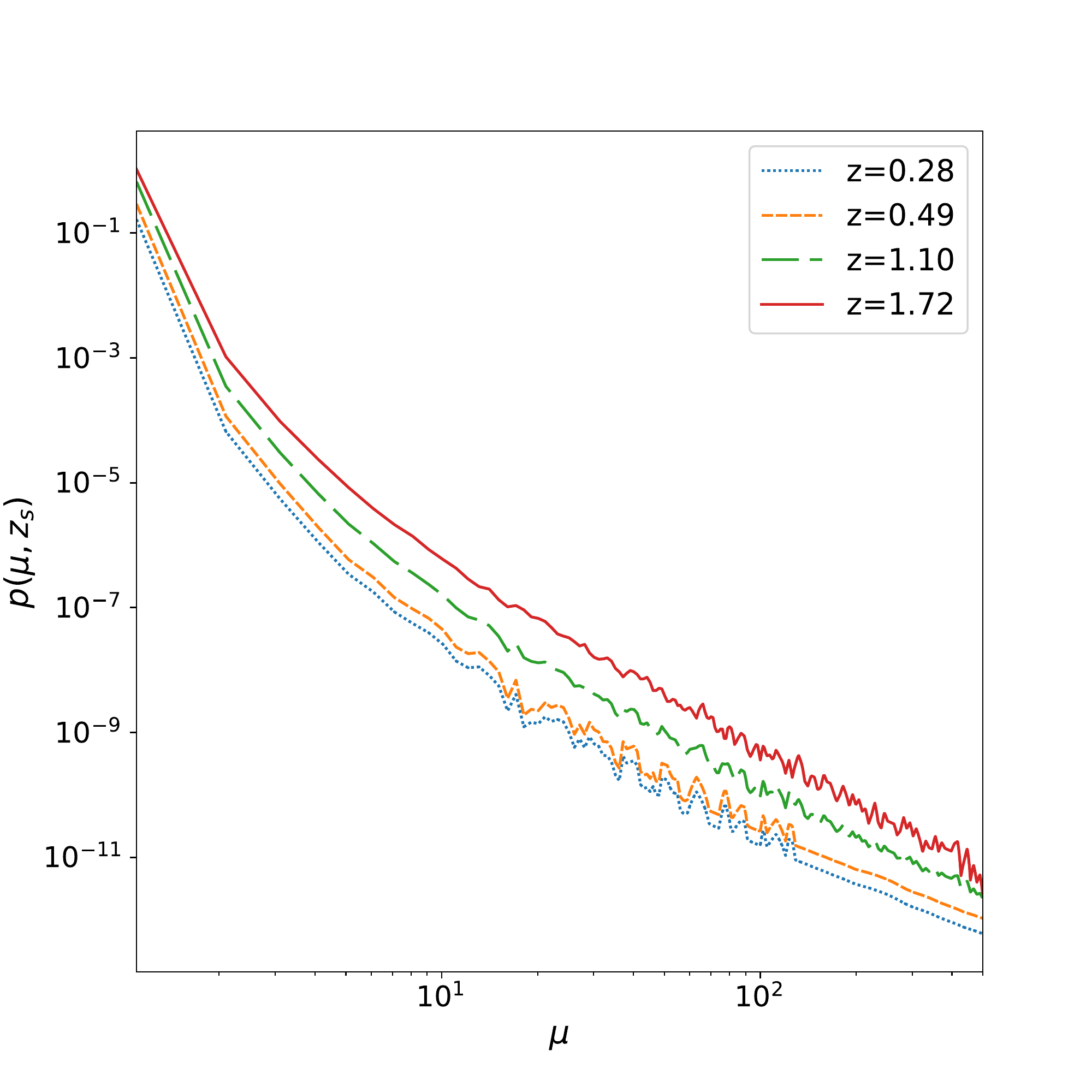}
    \caption{Magnification PDF for different values of source redshift: $z_{\rm s}=0.28$ (light blue, \textcolor{black}{dotted}), $z_{\rm s}=0.49$ (orange, \textcolor{black}{short-dashed}), $z_{\rm s}=1.10$ (green, \textcolor{black}{long-dashed}), $z_{\rm s}=1.72$ (red, \textcolor{black}{solid}).}
    \label{fig:dpdmu_taka}
\end{figure}


\section{Methods}

In this section we illustrate how we build our model for lens and source distributions. 

\label{sec:3}
\subsection{Lens distribution}
\label{sec:lens}
We  model the number density of galaxies (lenses) as a function of redshift and of the velocity dispersion $\sigma_{\rm v}$. 
One can show that if the evolution of sources is neglected, $\tau$ reduces to \citep[see, e.g.,][]{Cusin:19aa,Cusin:21aa}
\be\label{e:simp1}
\tau(\mu,z_s) \simeq \frac{0.001}{(\mu-1)^2}\left(\frac{H_0\chi(z_s)}{c}\right)^3\left[\frac{Nc^3}{10^9H_0^3}\frac{\langle\si_v^4\rangle}{c^4\times 5\times 10^{-14}}\right] \,,
\ee
where $N$ is the present galaxy density, 
\be\label{e:Nz0}
N= \int_0^\infty d\si_v n(\si_v,z=0) \,,
\ee
 and $\langle\si_{\rm v}^4\rangle$ a mean of the velocity dispersion to power 4.
Crude estimates for $N$ and $\si_{\rm v}$ are 
\be
N = 10^{9}\frac{H_0^3}{c^3} \,, \qquad \langle\si_{\rm v}^4\rangle =(150~\rm{km/s})^4 .
\ee
Then the PDF magnification reads
\begin{align}
    p(\mu, z_{\rm s})=\frac{2p_{1}(z_{\rm s})}{(\mu -1)^3}\exp{\left( \frac{p_{1}(z_{\rm s})}{(\mu-1)^2}\right)}, \,\,\,\, p_{1}(z_{\rm s})=0.001\left(\frac{H_{0}\chi(z_{\rm s})}{c}\right).
\end{align}
\textcolor{black}{\indent In the present work, we adopt this simplified model for the lens distribution at low redshift ($z<1$), which was validated by \citealt{Cusin:21aa}. Indeed, they show that such a method provides  results for the optical depth in good agreement with the one obtained  considering a more realistic distribution of lenses, which evolves with redshift (fractional deviations of a few percent). For completeness, we would also like to highlight an earlier work of \citealt{Shan20:aa}, where they used a full-sky multi-sphere ray-tracing code (developed by \citealt{Wei:18aa}) to calculate the GW lensing magnification at low redshift. Yet, for the purpose of our work, the analytic model of \citealt{Cusin:19aa,Cusin:21aa} is appropriate.}

We add to this strong lensing PDF the contribution of weak lensing due to the gravitational potential of the Large Scale Structure, which allows us to build a more realistic lens distribution. To do so, we use results for the magnification probability density of \citealt{Takahashi:11aa} (for $z\geq 1$), which reconstruct the path of light through inhomogeneous clumps of matter in the Universe via high-resolution ray-tracing approximation. We interpolate their results\footnote{The probability densities from \citealt{Takahashi:11aa} are available on this website \url{http://cosmo.phys.hirosaki-u.ac.jp/takahasi/raytracing/open_data/}} for the redshift values we want to study. In their simulations they used the box size of $50 h^{-1}$ Mpc with $10243$ particles, the mean particle separation of $50h^{-1}$ kpc, and the softening length of $2h^{-1}$ kpc.

The overall magnification PDF that we obtain is shown in Figure \ref{fig:dpdmu_taka}, for some selected values of $z_{\rm s}$. From this plot, we see that this function in general increases with the source redshift, which is reasonable since for bigger $z_{\rm s}$ we expect more foreground lenses between the source and the observer. Furthermore, over the magnification interval $1\leq\mu\leq 500$, the function decreases very steeply, showing a 10 order-of-magnitude lowering, which shows how higher values of the magnification are generally suppressed in favor of lower values.

\subsection{Source population}
\begin{figure}
    \centering
    \includegraphics[width=0.5\textwidth] {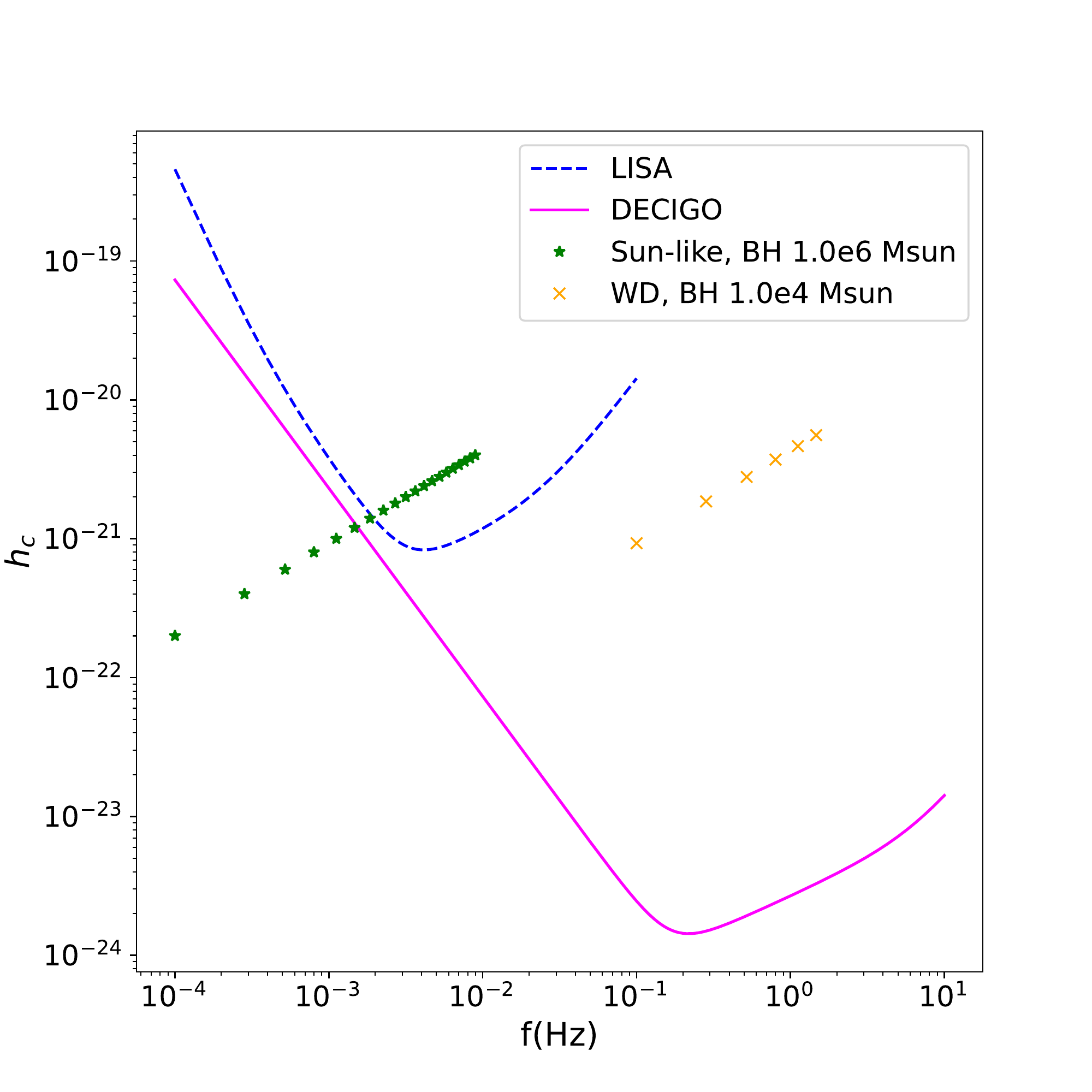}
    \caption{Sensitivity curves for LISA (blue, \textcolor{black}{dashed}) and DECIGO (magenta, \textcolor{black}{solid}). \textcolor{black}{On top of this, we also show the peak strain for two typical TDEs: Sun-like star disrupted by a $10^{6}\text{M}_{\odot}$ BH (green stars) and and WD disrupted by a $10^{4}\text{M}_{\odot}$ BH (orange cross), at a distance of 16 Mpc. The $\beta$ parameter is increasing from left to right (cf. Equation \ref{eq:beta_max}).}}
    \label{fig:curves}
\end{figure}
Before illustrating how we built the source population, we briefly recall the main formulas for the TDE gravitational emission.\\
\indent The GWs associated with the disruption of a star can be approximated as a monochromatic burst, with strain and frequency given by \citep[see, e.g.,][]{Toscani:22aa}
\begin{align}
     h_{\rm gw}& \approx 2\times 10^{-22}\beta\times\left(\frac{\text{M}_*}{\text{M}_\odot} \right)^{4/3}\times\left(\frac{M_{\rm bh}}{{10^{6}\text{M}_{\odot}}}\right)^{2/3}\times\nonumber \\
     &\left(\frac{R_*}{\text{R}_{\odot}}\right)^{-1}\times\left(\frac{\chi}{16\,\text{Mpc}}\right)^{-1},\\
    f_{\rm gw}& \approx\beta^{3/2}\times 10^{-4}\,\text{Hz}\times\left(\frac{M_*}{\text{M}_{\odot}}\right)^{1/2}\times\left(\frac{R_{*}}{\text{R}_{\odot}}\right)^{-3/2},
\end{align}
where $M_*$ and $R_*$ are the stellar mass and radius, $M_{\rm bh}$ the BH mass, $\text{M}_{\odot}$ and $\text{R}_{\odot}$ the solar mass and radius, and $\beta=R_{\rm t}/R_{\rm p}$ is the penetration factor, i.e. the ratio between the maximum distance from the BH to have a full disruption, a.k.a. the \textit{tidal radius}
\begin{align}
    R_{\rm t}\approx 7\times 10^{12}\,\text{cm}\times \left(\frac{R_{*}}{R_\odot}\right)\times\left(\frac{M_{\rm bh}}{10^6\text{M}_{\odot}}\right)^{1/3}\times\left(\frac{M_*}{\text{M}_\odot}\right)^{-1/3},
\end{align}
and the stellar pericenter $R_{\rm p}$. Requiring $R_{\rm p}$ bigger than BH Schwarschild radius, we get the following limits for $\beta$
\begin{align}
    1\lesssim\beta\lesssim\beta_{\rm max}\approx 20\times\left(\frac{R_*}{\text{R}_{\odot}}\right)\times\left(\frac{M_{*}}{\text{M}_\odot}\right)^{-1/3}\times\left(\frac{M_{\rm bh}}{10^{6}\text{M}_{\odot}}\right)^{-2/3}.
    \label{eq:beta_max}
\end{align}
\indent The SNR $\rho$ for such a signal can be written as (\citealt{Pfister:21aa}, see also Appendix \ref{app:snr})
\begin{align}
    \rho&=\frac{h_{\rm gw}}{h_{\rm c}(f_{\rm gw}/(1+z))}=\nonumber\\
    &=\beta \times 2 \times 10^{-22}\times \left( \frac{M_{*}}{M_\odot}\right)^{4/3}\times \left(\frac{M_ {\rm bh}}{10^6\text{M}_{\odot}}\right)^{2/3}\times\nonumber\\
    &\left( \frac{R_{*}}{\text{R}_{\odot}}\right)^{-1} \times\left(\frac{\chi}{16~\text{Mpc}}\right)^{-1}\times \frac{1}{h_{\rm c}(f_{\rm gw}/(1+z))},
    \label{eq:snr}
\end{align}
where $h_{\rm c}$ is the characteristic noise of the instrument \citep[][]{Moore:15aa}. Throughout this work, we consider the sensitivity curves of LISA \citep[]{LISAsr:18aa} and DECIGO \citep{Sato:17aa}, illustrated in Figure \ref{fig:curves}, \textcolor{black}{where we also plot the peak strain for two typical TDEs.}\footnote{\textcolor{black}{Since the TDE gravitational signal is a monochromatic burst, its Fourier transform can be written as $\tilde{h}\approx h_{\rm gw}/f_{\rm gw}$ and thus we have $h_{\rm c}\approx h_{\rm gw}$.}}\\
\indent In this study, when considering MS stars, we adopt the following approximated scaling relation \citep{Kippenhan:90aa}
\begin{align}
    \frac{M_{\rm ms}}{M_{\odot}}\approx \frac{R_{\rm ms}}{\text{R}_{\odot}},
    \label{eq:mass_radius_ms}
\end{align}
while for the WD case we assume fixed values for the mass and radius, $M_{\rm wd}=0.5\text{M}_{\odot}$, $R_{\rm wd}=0.01\text{R}_{\odot}$.\\
\indent Note that, from this point forward, we use $M_{\rm bh} \equiv M_{\bullet}$, when referring to massive BHs residing in galaxy cores, while we write $M_{\rm bh} \equiv M_{\rm h}$ when referring to IMBHs located in GCs.

\subsubsection{Nuclear TDEs}
We build the population of MS stars tidally disrupted by massive BHs residing in galaxy cores following the same steps as in \citealt{Toscani:20aa}, 
\begin{align}
    \frac{d\mathcal{N}^{\rm ms}}{dzdM_\bullet dM_{\star}d\beta}=\frac{4\pi c\chi(z)^2}{H_0E(z)}\Phi(M_\bullet)\psi(\beta)\phi(M_*)\frac{\Gamma(M_{\bullet})}{(1+z)}T,
    \label{eq:mspop}
\end{align}
where we have the following terms:
\begin{enumerate}
\item the comoving volume term ${4\pi c\chi(z)^2}/{H_0E(z)}$;
\item the distribution of nuclear massive BHs that we build from a Schechter mass function with z-dependence \citep[see]
[]{Gabasch:06aa}, expressed in terms of $M_{\bullet}$ using the Faber-Jackson relation \citep[][]{Faber:76aa} and the $M_{\bullet}-\sigma$ relation \citep[][]{McConnell:13aa}, as done in \citealt{Toscani:20aa},
\begin{align}
    \red{\Phi(M_\bullet)}=&\frac{0.003\text{Mpc}^{-3}}{(1+z)^{0.48}10^8\text{M}_{\odot}}\times \left( \frac{M_\bullet}{10^8\text{M}_{\odot}}\right)^{-1.24}\\\nonumber
    &\times\exp{\left(-\frac{0.59}{(1+z)^{0.7}} \left(\frac{M_\bullet}{10^8\text{M}_{\odot}}\right)^{0.7}\right)};
\end{align}
\item the normalized distribution for $\beta$ \citep[][]{Stone:16aa}
\begin{align}
   \psi(\beta)= \frac{\beta_{\rm max}(M_{\bullet}, M_*)}{\beta^2(\beta(M_{\bullet}, M_{*\rm max})-1)};
\end{align}
\item the normalized Salpeter initial stellar mass function \citep{Salpeter:55aa}
\begin{align}
    \phi(M_*)=\frac{1.35}{M^{-1.35}_{*\rm min}-M^{-1.35}_{*\rm max}}M_*^{-2.35};
\end{align}
\item the galaxy rate for nuclear TDEs \citep[see, e.g,][]{Stone:16aa}
\begin{align}
    \Gamma(M_\bullet)=2.9\times 10^{-5}\,\text{/(yr gal)}\left( \frac{M_{\bullet}}{10^8\text{M}_{\odot}}\right)^{-0.404};
\end{align}
\item the observation time $T$, that we take equal to the lifetime of the mission. Here we assume, both for LISA and DECIGO, $T=10 \,\text{yr}$. 

\end{enumerate}
\begin{figure*}
  \centering
  \includegraphics[width=0.48\textwidth]{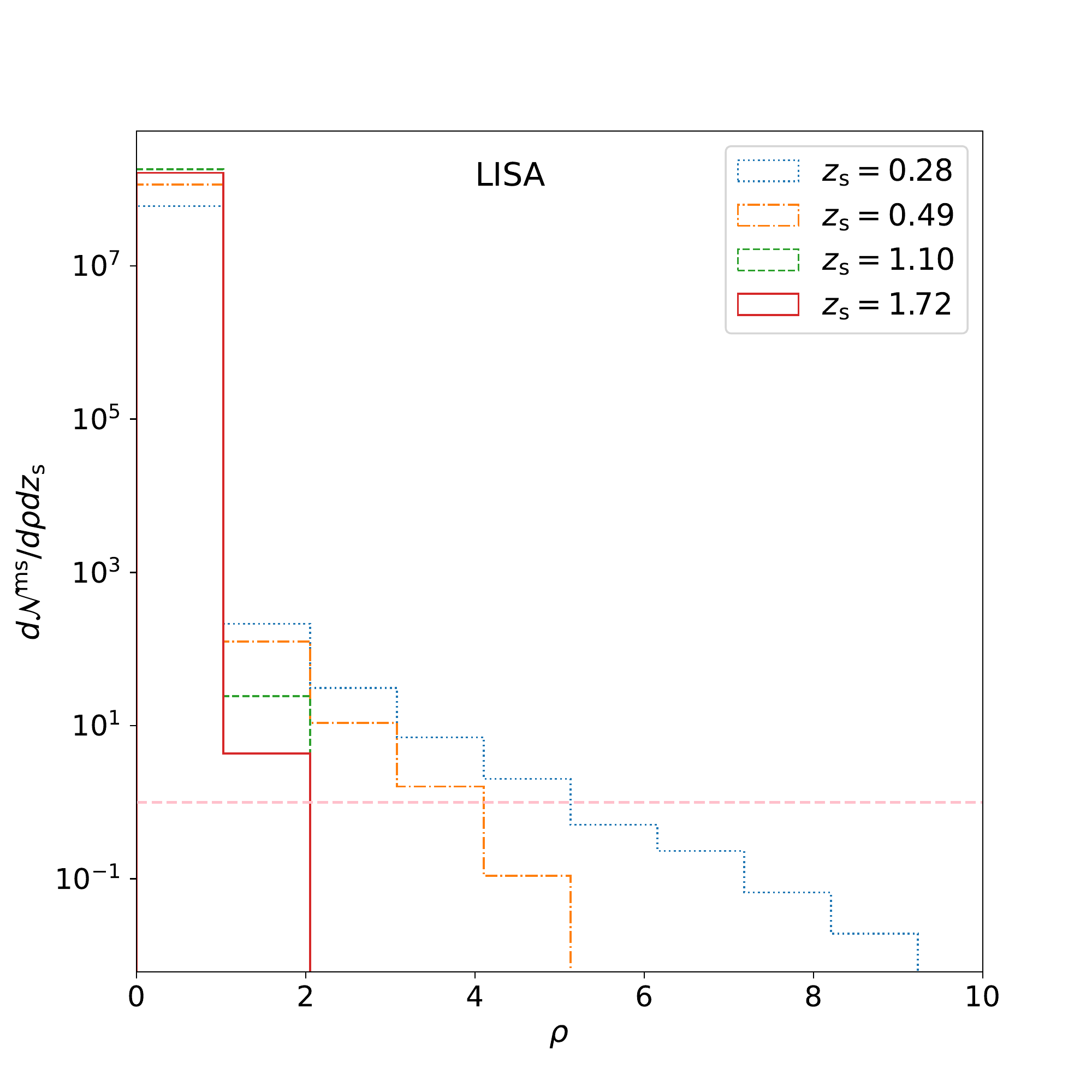}
\includegraphics[width=0.49\textwidth]{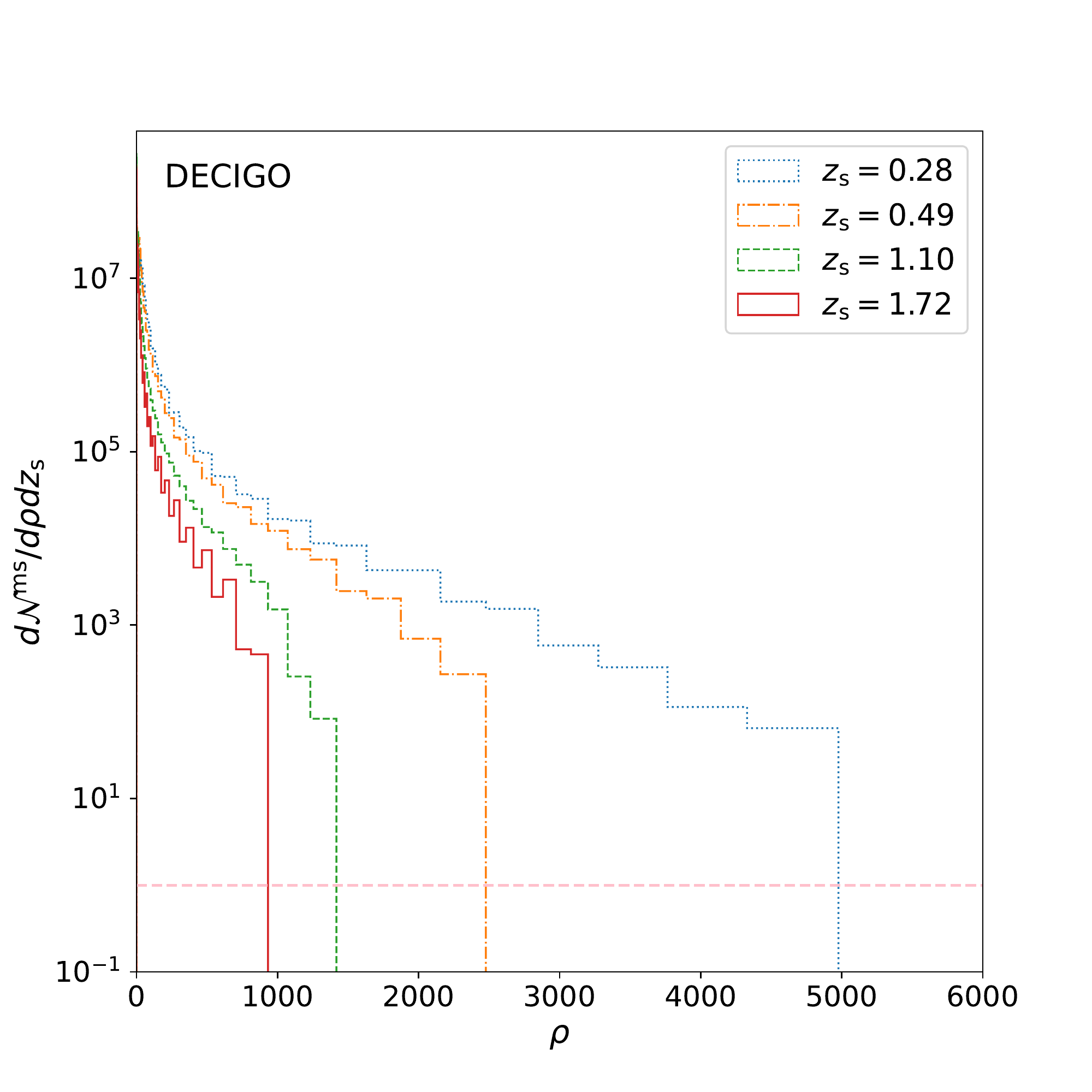}
\caption{Number of nuclear TDEs per bin of SNR and bin of source redshift as a function of SNR $\rho$. \textcolor{black}{The different redshift bins are represented in the following way: $z_{\rm s}=0.28$ blue dotted line, $z_{\rm s}=0.49$ orange dot-dashed line, $z_{\rm s}=1.10$ green dashed line, $z_{\rm s}=1.72$ red solid line.} On the left panel we consider LISA, on the right panel we consider DECIGO. The pink horizontal dashed line represents 1 TDE per SNR and redshift bin in 10 years.} 
  \label{fig:dndzdrho_ms}
\end{figure*}
\subsubsection{Globular TDEs}
We build the population of WDs tidally disrupted by IMBHs located in GCs in a similar way as done in \citealt{Toscani:20aa},
\begin{align}
 \frac{d\mathcal{N}^{\rm wd}}{dzdM_\bullet d\beta}=\frac{4\pi c\chi(z)^2}{H_0E(z)}\Phi(M_\bullet)N^{\rm gc}_{\rm gal}(M_{\bullet})\psi(\beta)\frac{\Pi(M_{\rm h}
,M_{\rm wd})}{(1+z)}T,
 \end{align}
 where we have the following terms:
 \begin{enumerate}
     \item the comoving volume term ${4\pi c\chi(z)^2}/{H_0E(z)}$;
     \item the distribution of nuclear massive BHs, $\Phi(M_{\bullet})$;
     \item a scaling relation between the number of GCs per galaxy and the mass of the BH in the core \citep[][]{Harris:11aa,Burkert:10aa}
\begin{align}
    N^{\rm gc}_{\rm gal}= \frac{M_{\bullet}}{4.07\times 10^5 M_{\sun}};
\end{align}
\item the rate of globular TDEs per GCs \citep[][]{Baumgardt:04aa}
\begin{align}
     \Pi\sim \textcolor{black}{60}\text{Myr}^{-1}&\times\left(\frac{R_{\rm wd}}{\text{R}_{\sun}}\right)^{4/9}\times\left( \frac{M_{\rm wd}}{\text{M}_{\sun}}\right)^{-95/54}\times\nonumber\\
    &\times \left(\frac{M_{\rm h}}{10^3\text{M}_{\sun}}\right)^{61/27}\times\left( \frac{n_{\rm c}}{\text{pc}^{-3}}\right)^{-7/6}\times\left(\frac{r_{\rm c}}{\text{1pc}} \right)^{-49/9},
\end{align}
where we take the GC core density equal to $n_{\rm c}=10^{5}\,\text{pc}^{-3}$ and the GC core radius equal to $r_{\rm c}=0.5\,\text{pc}$. We remind that we assume the WD mass and radius to be fixed, $M_{\rm wd}=0.5\text{M}_{\odot}$, $R_{\rm wd}=0.01\text{R}_{\odot}$.
 \end{enumerate} 
We assume that the mass distribution of IMBHs
is a $\delta$ function at a fixed value of $M_{\rm h}$. In particular, we will build two populations, one with $M_{\rm h}=10^3\text{M}_{\odot}$, the other with $M_{\rm h}=10^4\text{M}_{\odot}$.

 \begin{figure*}
    \centering
    \includegraphics[width=0.48\textwidth]{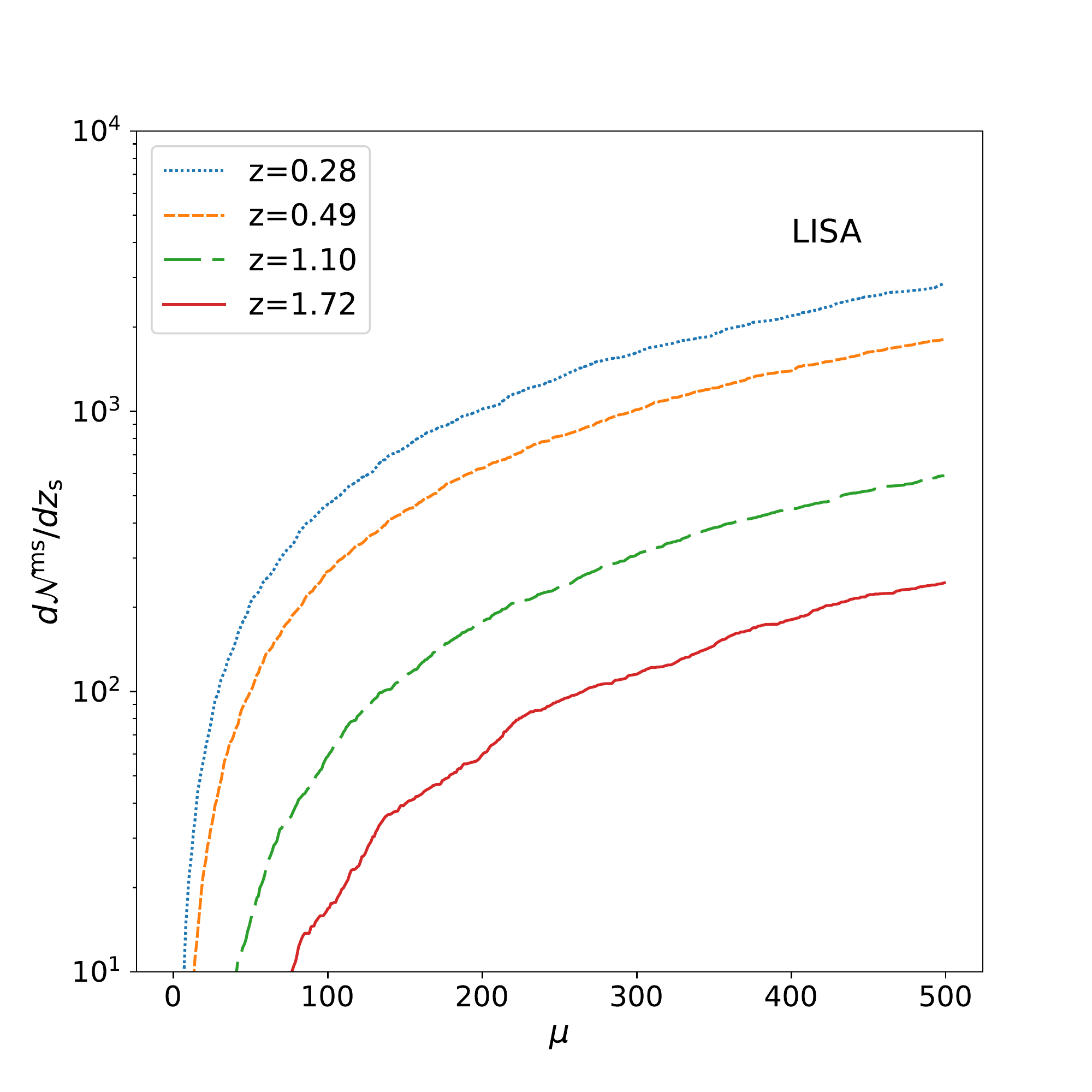}
  \includegraphics[width=0.49\textwidth]{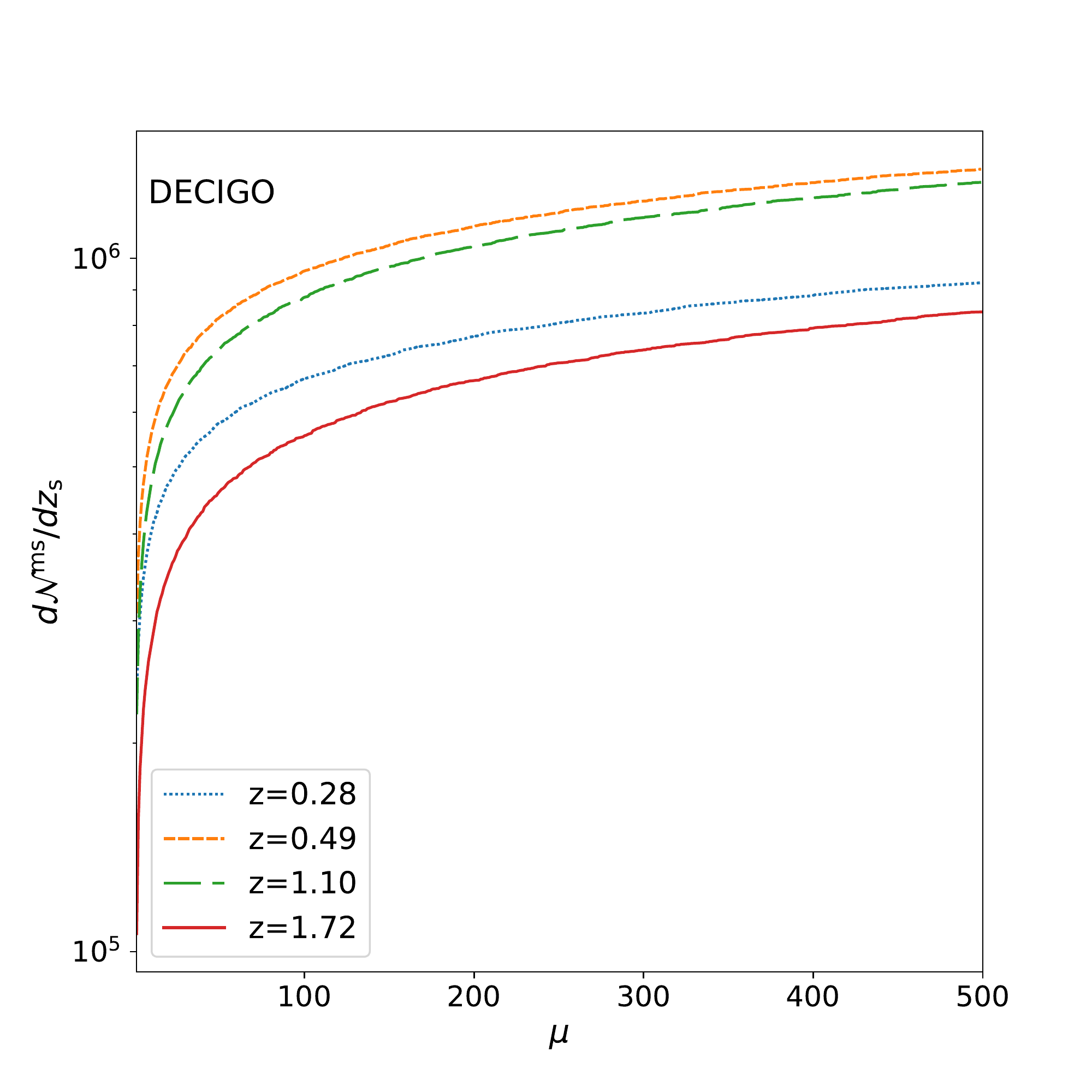}
  \caption{Number of visible (i.e. above the threshold $\rho_{\rm lim}/\sqrt{\mu}$) nuclear TDEs if the magnification is $\mu$. On the left panel we consider LISA, on the right panel we consider DECIGO. \textcolor{black}{The different redshift bins are represented in the following way: $z_{\rm s}=0.28$ blue dotted line, $z_{\rm s}=0.49$ orange short-dashed line, $z_{\rm s}=1.10$ green long-dashed line, $z_{\rm s}=1.72$ red solid line.} } 
    \label{fig:dndz_ms}
  \end{figure*}
\subsubsection{Range of parameters for the TDE populations}
To derive the total number of observed TDEs in presence of magnification, $\mathcal{N}^{\rm obs}$, we start by building the nuclear and globular TDE populations according to the aforementioned description. In particular we choose the following ranges: 
\begin{itemize}
    \item for the source redshift we take $z_{\rm s}\in [0.001,2]$, where the minimum value of redshift corresponds to $\approx 20 \text{Mpc}$ (average distance of the Virgo Cluster), while the maximum value corresponds to the redshift after which the GW emission from a TDE population is negligible \citep[][]{Toscani:20aa}.
   \item for the central BH mass we take $M_{\bullet}\in[10^{4}\text{M}_{\odot},10^{9}\text{M}_\odot]$, thus considering both dwarf and large galaxies;
   \item for the stellar mass we take $M_*\in [1 \text{M}_{\odot},100\text{M}_{\odot}]$ for the MS star case, hence young stellar population, while for the WD case we assume fixed mass and radius $M_{\rm wd}=0.5\text{M}_{\odot}$, $R_{\rm wd}=0.01\text{R}_{\odot}$.
   \item for the penetration factor we take $\beta\in[1,\beta_{\rm max}]$, where the formula for $\beta_{\rm max}$, which in general will depend on the BH and star mass, is illustrated in Equation \eqref{eq:beta_max}. 
\end{itemize}
\begin{figure*}
  \centering
  \includegraphics[width=0.48\textwidth]{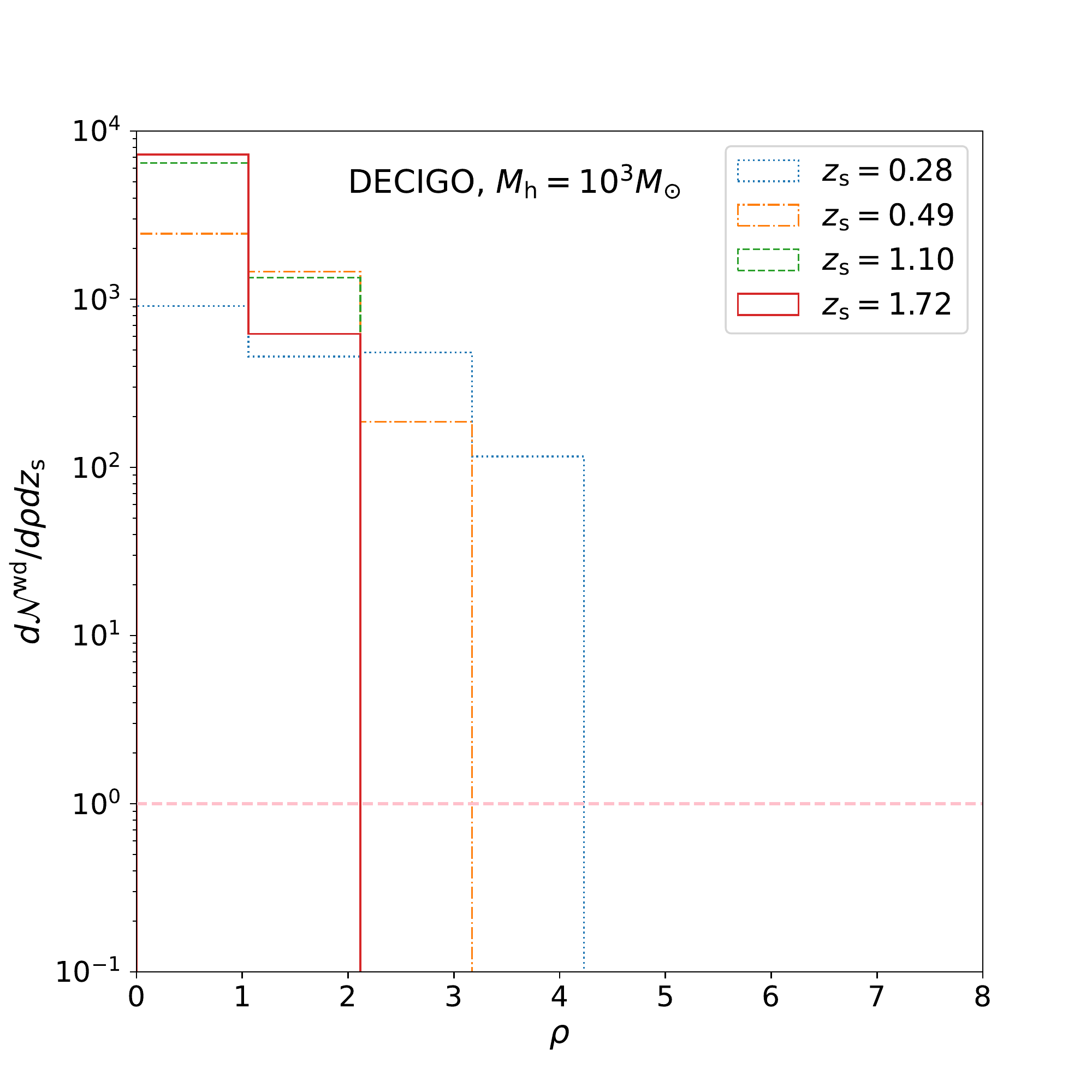}
\includegraphics[width=0.49\textwidth]{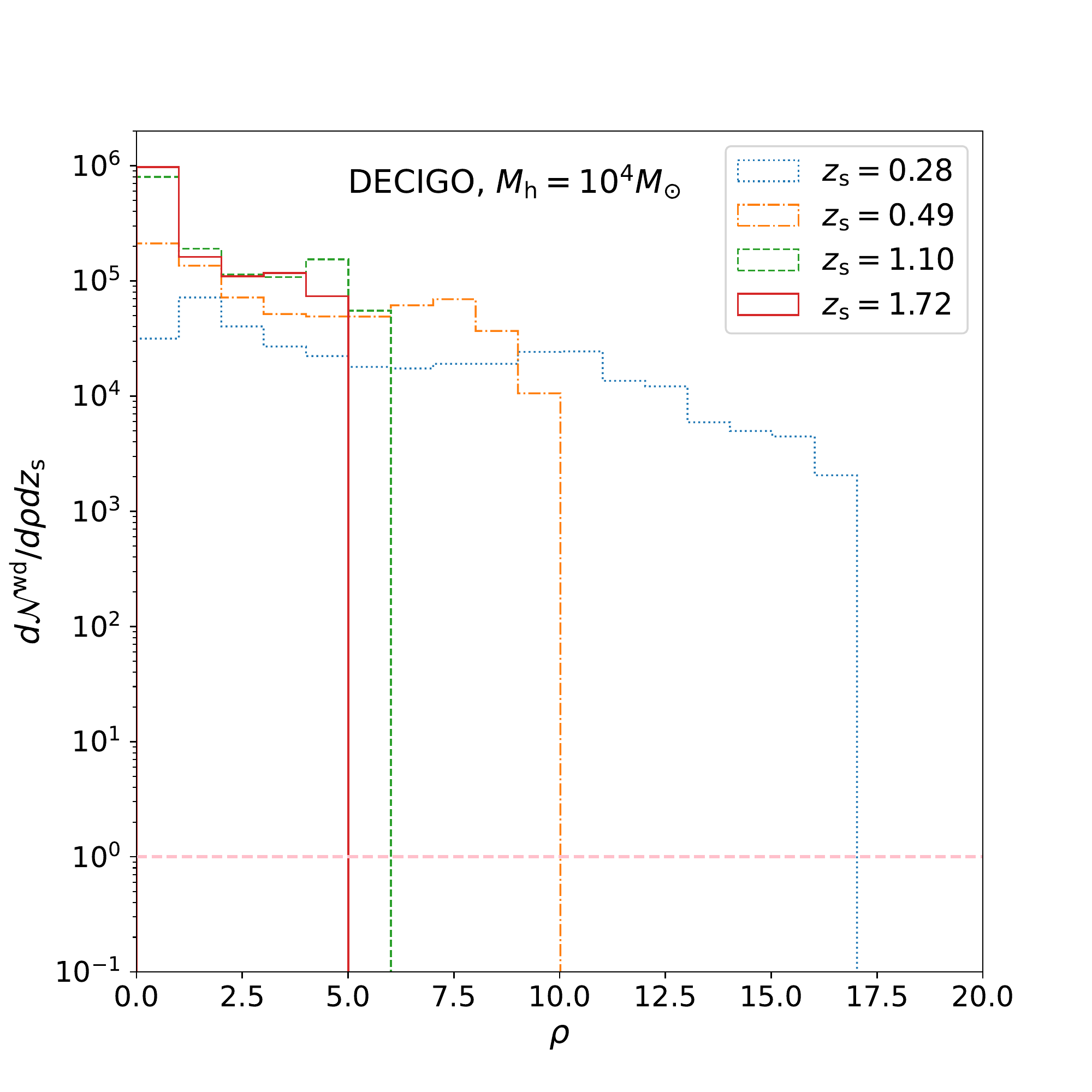}
\caption{Number of globular TDEs per bin of SNR and bin of source redshift. \textcolor{black}{The different redshift bins are represented in the following way: $z_{\rm s}=0.28$ blue dotted line, $z_{\rm s}=0.49$ orange dot-dashed line, $z_{\rm s}=1.10$ green dashed line, $z_{\rm s}=1.72$ red solid line.} On the left panel we consider $M_{\rm h}=10^{3}\text{M}_{\odot}$, on the right $M_{\rm h}=10^{4}\text{M}_{\odot}$. We assume that the interferometer is DECIGO. The pink horizontal dashed line represents 1 TDE per SNR and redshift bin in 10 years.} 
  \label{fig:dndzdrho_wd}
\end{figure*}
 
\begin{figure*}
    \centering
    \includegraphics[width=0.48\textwidth]{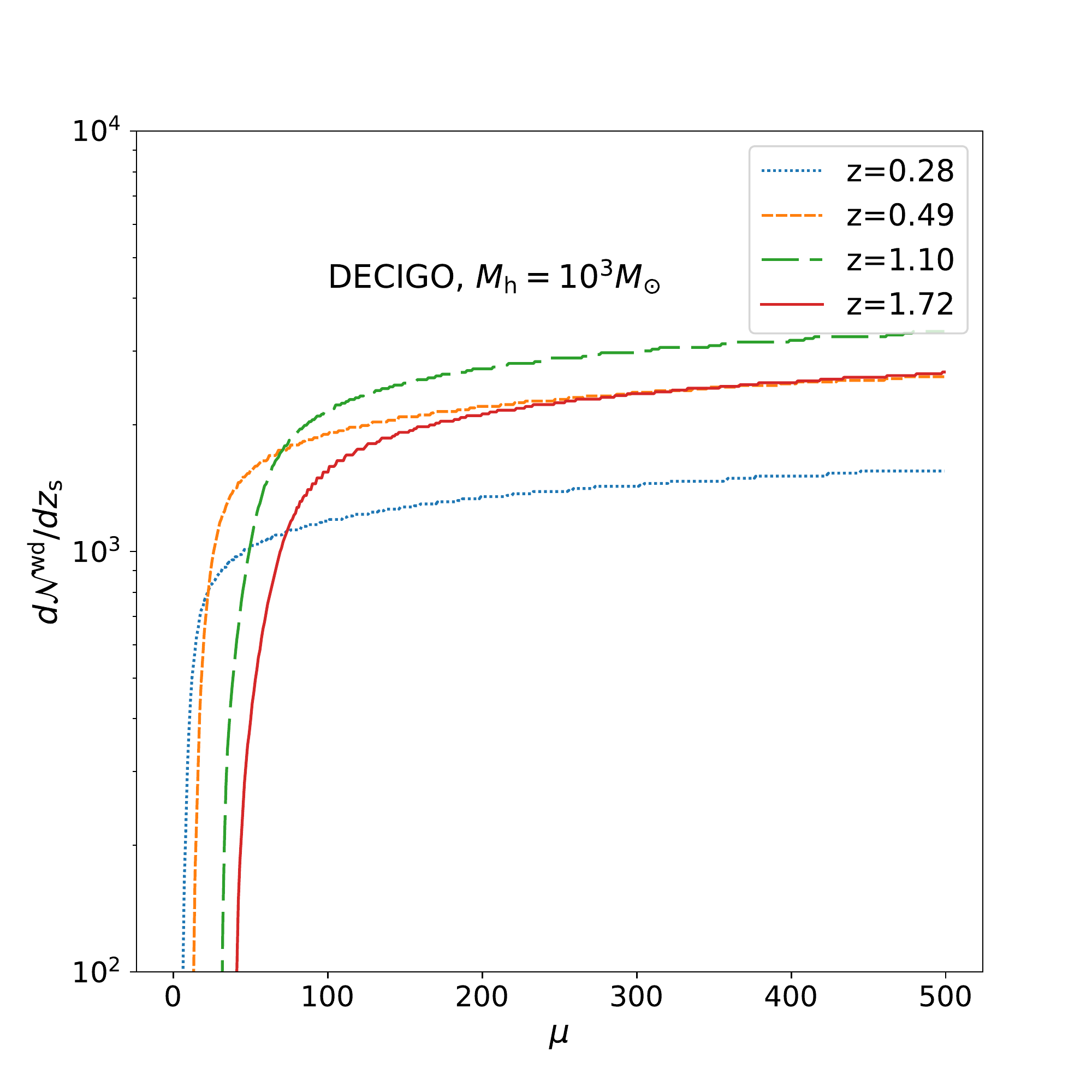}
  \includegraphics[width=0.49\textwidth]{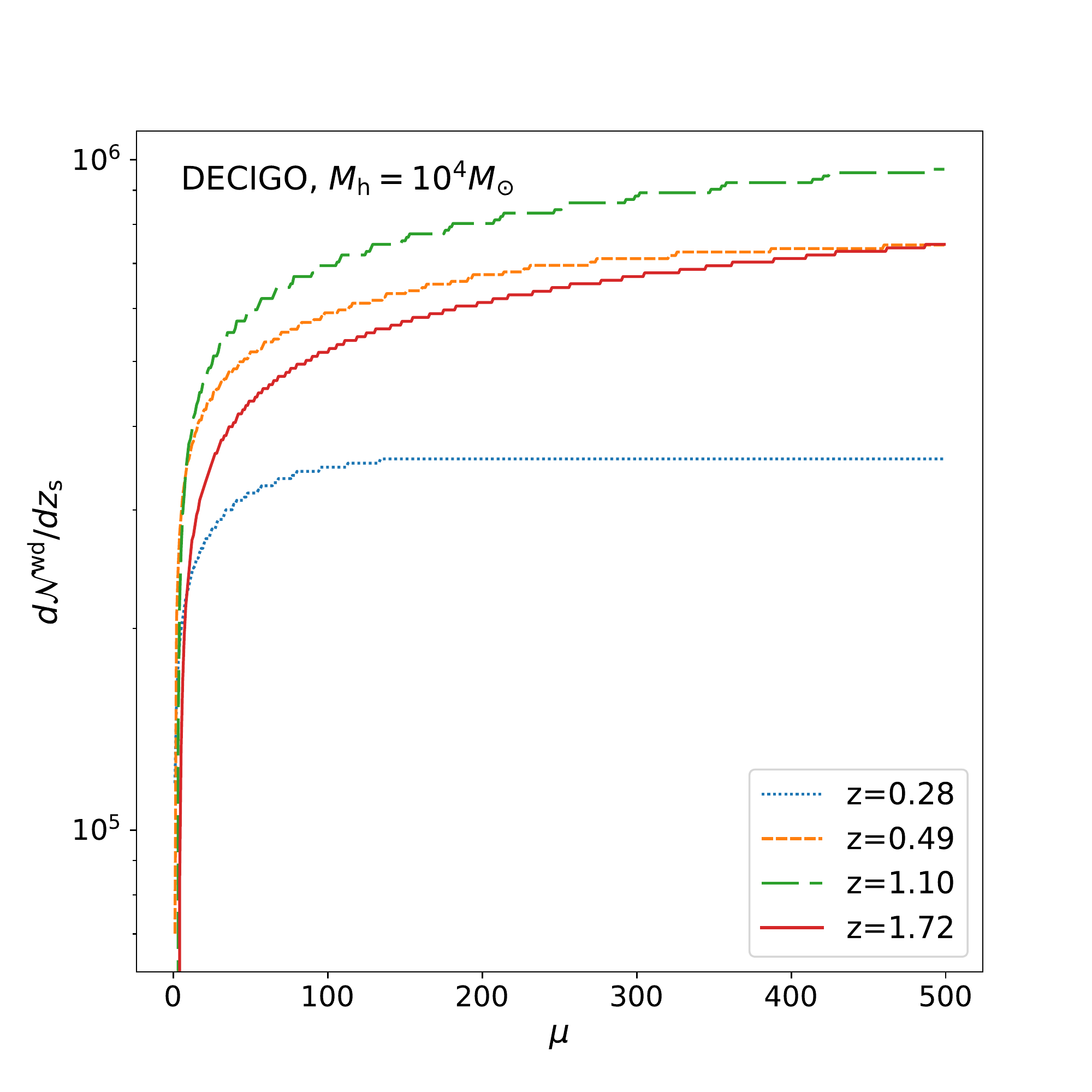}
  \caption{Number of visible (i.e. above the threshold $\rho_{\rm lim}/\sqrt{\mu}$) globular TDEs if the magnification is $\mu$. On the left panel we consider $M_{\rm h}=10^{3}\text{M}_{\odot}$, on the right panel we consider $M_{\rm h}=10^{4}\text{M}_{\odot}$. \textcolor{black}{The different redshift bins are represented in the following way: $z_{\rm s}=0.28$ blue dotted line, $z_{\rm s}=0.49$ orange short-dashed line, $z_{\rm s}=1.10$ green long-dashed line, $z_{\rm s}=1.72$ red solid line.} } 
    \label{fig:dndz_wd}
  \end{figure*}

\section{Results}
\label{sec:4}
Once we have built the two TDE populations, we can study which is the probability that they are lensed by a foreground population of lenses, using the framework presented in Section \ref{sec:lens}. We illustrate the main results of our lensing study in the following, distinguishing between the case of nuclear and globular TDEs.

\subsection{Strong lensing of GWs from nuclear TDEs}
In Figure \ref{fig:dndzdrho_ms} we show the number of MS stars tidally disrupted by massive BH per bin of SNR and $z_{\rm s}$, $d{\mathcal{N}^{\rm ms}}/d\rho d z_{\rm_{s}}$ (cf. Equation \ref{eq:dn_dz}). On the left side we calculate the SNR considering LISA, on the right side DECIGO. Each colour represents the number of TDEs in a redshift bin $z_{\rm s}\pm 0.068$, and in particular we show selected values of $z_{\rm s}$: 0.28 (blue), 0.49 (orange), 1.10 (green), 1.72 (red). In general, we see that $d{\mathcal{N}^{\rm ms}}/d\rho d z_{\rm_{s}}$ diminishes for higher values of $\rho$ as expected. In addition to this, the maximum SNR decreases for higher redshift. As for the minimum SNR, this is always $\approx 0$, which is a reasonable result since for each redshift bin we have TDEs below the instruments sensitivity curves.\\
\indent In Figure \ref{fig:dndz_ms}, we show the number of visible (i.e. above the threshold ${\rho_{\rm lim}}/{\sqrt{\mu}}$) nuclear TDEs if the magnification is $\mu$, $d\mathcal{N}^{\rm ms}(\mu,z_{\rm s})/dz_{\rm s}$, calculated through Equation \eqref{eq:dn_dz}. The layout and colour are the same as in Figure \ref{fig:dndzdrho_ms}. From these plots, we see that while for LISA $d\mathcal{N}^{\rm ms}(\mu,z_{\rm s})/dz_{\rm s}$ decreases for higher values of redshift, the same quantity for DECIGO first increases for higher values of $z_{\rm s}$, than it starts to lower again. This behavior can be explained in the following way. Since DECIGO has a better sensitivity than LISA, there are two opposite trends that interplay between each other: i) the total number of TDEs increases for higher values of redshift (volume effect), ii) the number of visible TDEs decreases for high values of redshift (SNR limitation). In the case of LISA however, 
which presents a worse sensitivity to TDEs, the ii) effect always prevails.
In other words, LISA is always SNR limited and the very few detectable events decrease rapidly with redshift.\\
\indent Finally, we have all the ingredients to calculate the total number of observed TDEs in presence of magnification, $\mathcal{N}^{\rm obs}$, through Equation \eqref{eq:dn_obs}. Restricting to magnification $\mu>2$ (i.e. focusing on the stronger lensed image, cf. Section \ref{sec:2} and Appendix \ref{sec:strong}) , we find that for LISA the number of lensed-magnified TDEs is 0, while for DECIGO we expect the detection of $\sim$135 magnified TDEs ($\mu>2$). Yet, this number decreases quite rapidly as we increase the magnification threshold: it reduces to $\sim$18 for $\mu>3$, $\sim$6 for $\mu>4$, $\sim$3 for $\mu>5$ and quickly drops for higher magnification ($\sim 0$ for $\mu >10$). This fast drop is in agreement with the steep decreasing presented by the magnification PDF illustrated in Figure \ref{fig:dpdmu_taka}.

\subsection{Strong lensing of GWs from globular TDEs}
As for the case of globular TDEs, we consider two sub-populations: one where WDs are disrupted by IMBHs of mass $10^{3}\text{M}_{\odot}$, the other with an IMBH mass of $10^{4}\text{M}_{\odot}$. We assume that in each GC there is an IMBH.\\
\indent If we consider LISA, both these sub-populations of globular TDEs are below threshold and not even lensing can make part of these sources detectable.\\
\indent The situation is instead more interesting if we consider DECIGO. In Figure \ref{fig:dndzdrho_wd} we show the number of WDs tidally disrupted by IMBHs per bin of SNR and $z_{\rm s}$, $d{\mathcal{N}^{\rm wd}}/d\rho d z_{\rm_{s}}$. On the left side we consider $M_{\rm h}=10^{3}\text{M}_{\odot}$, while on the right $M_{\rm h}=10^{4}\text{M}_{\odot}$. The colour scheme is the same as previously described. In a similar way as for the nuclear TDE scenario, we note that: i) $d{\mathcal{N}^{\rm wd}}/d\rho d z_{\rm_{s}}$ shows a general decreasing trend while $\rho$ increases; ii) the maximum SNR decreases for for higher $z_{\rm s}$; iii) the minimum SNR is $\approx 0$.\\
\indent In Figure \ref{fig:dndz_wd}, we show the number of observable globular TDEs if the magnification is $\mu$, $d\mathcal{N}^{\rm wd}(\mu,z_{\rm s})/dz_{\rm s}$, calculated through Equation \ref{eq:dn_dz}.
The layout and colour are the same as in the previous plot. Also in this case, we note the interplay between the volume  effect against the SNR effect already presented in Figure \ref{fig:dndz_ms} for DECIGO.\\
Finally we have all the ingredients to calculate the number of observed magnified TDEs. For the case $M_{\rm h}=10^{3}\text{M}_{\odot}$, 
DECIGO will not observe any TDEs with $\mu >2$. Thus, DECIGO will not detect any TDEs from this population. As for the scenario $M_{\rm h}=10^{4}\text{M}_{\odot}$, the number of TDEs is $\sim$16 for $\mu>2$, $\sim$6 for $\mu>3$, $\sim$3 for $\mu>4$, $\sim$2 for $\mu>5$ and it quickly drops for higher magnification ($\sim 0$ for $\mu >10$).
\begin{figure}
    \centering
    \includegraphics[width=0.5\textwidth]{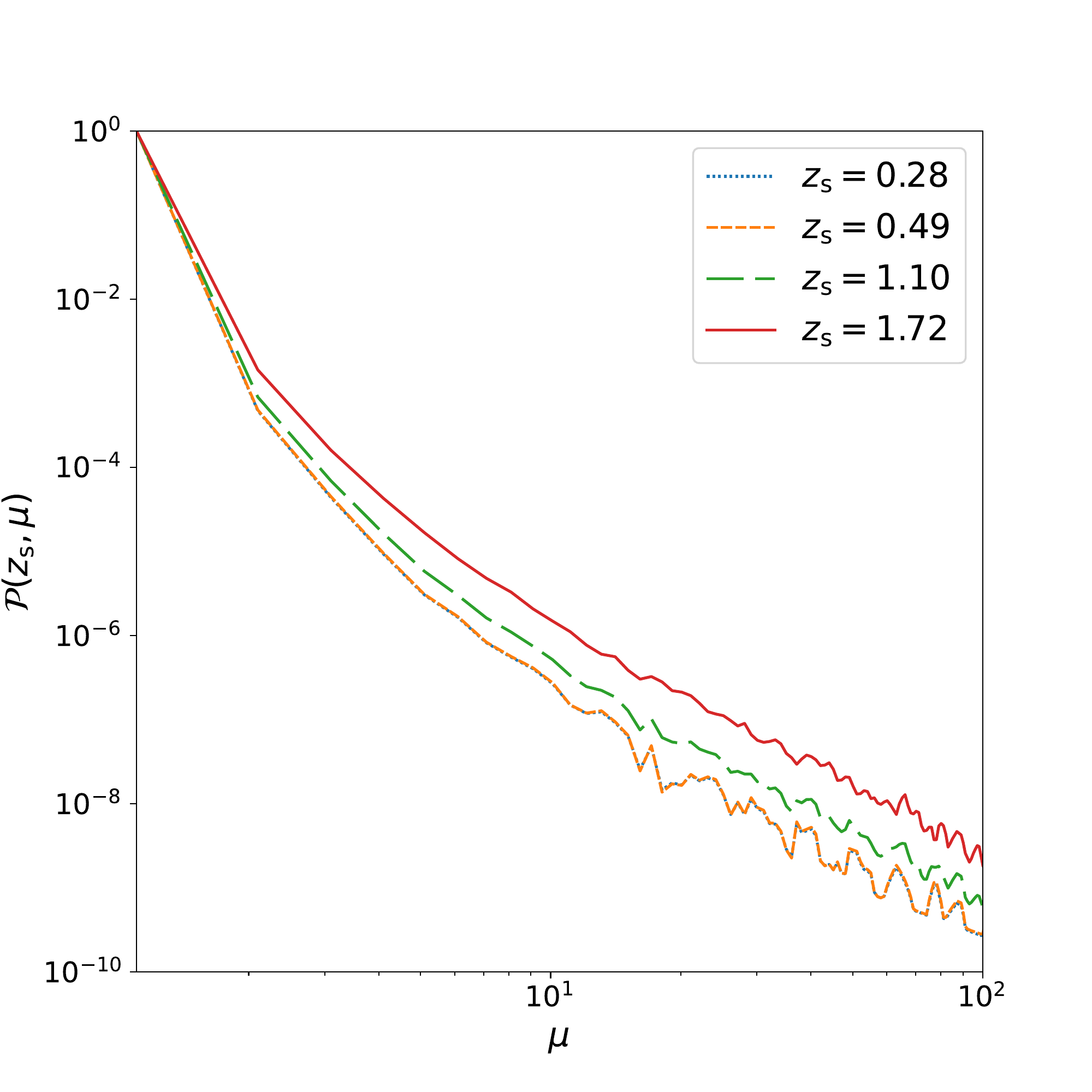}
    \caption{Probability that, if DECIGO detects a nuclear TDE from redshift $z_{\rm s}$, this is magnified more than $\mu$. \textcolor{black}{The different redshift bins are represented in the following way: $z_{\rm s}=0.28$ blue dotted line, $z_{\rm s}=0.49$ orange short-dashed line, $z_{\rm s}=1.10$ green long-dashed line, $z_{\rm s}=1.72$ red solid line.} }
    \label{fig:P_dec}
\end{figure}
\section{Discussion and conclusions}
\label{sec:5}
In this paper, we have investigated the effects of gravitational lensing of GW signals from
TDEs of MS stars disrupted by massive BHs in galaxy cores (nuclear TDEs) and from TDEs of WDs disrupted by IMBHs in GCs (globular TDEs). In order to follow a most realistic procedure as possible, we built the distribution of lenses following an hybrid approach. 
The lenses (galaxies) are modeled as SIS. To derive numerical results, we add the contribution of weak lensing from the Large Scale Structure using the results from ray-tracing simulations of \citet{Takahashi:11aa}, which include both weak
lensing contribution, and strong lensing tails in the magnification PDF.

For the TDE population, we follow similar steps as in \citet{Toscani:20aa}. Our work shows that, while LISA shall not be able to observe lensed-magnified TDEs, the situation will be different for an interferometer with a DECIGO-like sensitivity. While this interferometer will observe $\sim$16 magnified globular TDEs ($\mu>2$), \textcolor{black}{for the most promising scenario of nuclear TDEs it should detect $\sim$135 events with $\mu>2$ (78\% of which at redshift higher then 1)}. In Figure ~\ref{fig:P_dec} we show a summary plot illustrating, for this latter case, the probability that a TDE from  redshift $z_{\rm s}$ is magnified more than $\mu$ (see Equation \ref{eq:prob_mag}).
The probability that a TDE observed at redshift $z_{\rm s}=0.28$ would be magnified more than $\mu>2$ is $\sim$$10^{-4}$, and increases up to one order of magnitude if we go to higher redshift ($z_{\rm s}=1.72$). At fixed redshift bin, the probability to have higher magnification decreases steeply (roughly 9 to 10 orders of magnitudes in the interval $1\leq \mu\leq 100$), which justifies why the number of magnified TDEs drops rapidly as explained above.\\
\indent Our results point out that DECIGO will observe a non-negligible fraction ($\sim$$0.1\%$) of strongly lensed TDEs. Hence data analysis techniques need to be developed to be able to distinguish lensed TDEs from unlensed ones. This will be important to prevent a biased reconstruction of the parameters of the source. Lensed events will in fact have a (de-)magnified GW amplitude at the detector, which could bias the measurement of source parameters such as its distance.
Furthermore, we expect lensing to induce \textit{lensing selection effects} on the study of the TDE population, in analogy with what found for example in \cite{Cusin:21ab} for a population of supermassive black hole binaries visible by LISA. Indeed, a realistic GW detector has a finite sensitivity: magnified sources are on average easier to detect than de-magnified ones and this affects the distribution of lensing magnification of an observed source sample. These lensing selection effects, which should disappear in the limit of a perfect instrument, are then expected to introduce a bias on the reconstruction of the source parameters (e.g.~the luminosity distance), independent of the sample size. Hence, the characterisation of all the implications due to lensing, including selection effects due to the specifics of a given instrument, is
necessary to accurately infer
the source population astrophysical properties across cosmic time, but also to be able to use high-redshift GW sources as a new cosmological probe. 

We would like to remind that our study of lensing relies on the geometric optics approximation. We expect wave effects to become non-negligible in the mHz waveband only when dealing with diffusion off sub-galactic structures, see e.g.\,\citet{Takahashi:2003ix,PhysRevLett.80.1138,Takahashi:2016jom, Dolan:2007ut, Cusin:2018avf,  Cusin:2019rmt, Dalang:2021qhu}. Diffraction on sub-galactic scales makes lenses on those scales effectively transparent to GW in the LISA band \citep{Takahashi:2003ix}, in contrast with what happens for lensing of EM sources (see e.g.\,\citealt{Fleury_2015}) as the EM spectrum is at much lower wavelengths than any relevant astrophysical structure at cosmological scales.

We conclude by pointing out that TDEs may indeed constitute highly interesting multi-messenger sources, as they emit not only EM radiation, but also GWs and high-energy neutrinos.
The observation of both EM and GW signals from the same TDE, could in fact enable spectacular multi-messenger analyses which may well unveil new secrets on the intrinsic mechanisms behind these events.
Concretely for example, the GW signal would mark the moment of stellar disruption at the first pericenter passage, otherwise undetectable \citep{Rossi:20aa}. A measurement of the time delay between the GW signal and the subsequent EM signals would decisively help discriminating between EM emission mechanisms, currently highly debated \citep[][]{Bonnerot:20aa}.
We may compare this scenario with GW170817 which revolutionised our understanding of binary neutron star mergers \citep{LIGOScientific:2017vwq,LIGOScientific:2017zic,LIGOScientific:2018hze}, and triggered new tests of general relativity \citep{LIGOScientific:2018dkp} and new cosmological measurements \citep{LIGOScientific:2017adf}.
TDEs may well be used for similar measurements in the future: for example they could allow us to probe the expansion of the universe if the luminosity distance is extracted from the GW signals and the host galaxy is identified from the EM emission, in analogy to massive BH binary mergers with LISA \citep{Tamanini:2016zlh,LISACosmologyWorkingGroup:2019mwx} and double WD mergers with DECIGO \citep{Maselli:2019mzt}.

The observation of a multi-messenger lensed TDE would not only provide the data for the analyses outlined above, but the differences in the observed EM and GW lensed signals would yield unprecedented opportunities to study properties of both the source and the lens.
For example the EM radiation will always fall within the geometric optics approximation, while as mentioned above the GW signal could show signs of wave optics effects, which may then be used to infer additional information on the lens.
A detailed analysis of lensed EM emission from TDEs is currently missing in the literature and this will be the subject for a future work.

\section*{Acknowledgements}
 
\textcolor{black}{The authors are thankful to the referee Bin Hu, who provided useful comments that improve the quality of this work.} MT and NT \textcolor{black}{acknowledge support from an ANR Tremplin ERC Grant (ANR-20-ERC9-0006-01)}. The work of GC is supported by CNRS and by the Swiss National Science Foundation (Ambizione grant--\emph{Gravitational wave propagation in the clustered Universe}). EMR acknowledges that this project has received funding from the European Research Council (ERC) under the European Union’s Horizon 2020 research and innovation programme (Grant agreement No. 101002511 - VEGA P).

\section*{Data Availability}

The data underlying this article will be shared on reasonable request to the corresponding author.



\bibliographystyle{mnras}
\bibliography{bibliografia} 



\appendix
\section{Strong lensing in a nutshell}
In this Appendix we present details of the derivation of cross-section and optical depth for a SIS lens.  

\label{sec:strong}
\subsection{Basic strong lensing quantities}
A typical situation considered in gravitational lensing is the one illustrated in Figure \ref{fig:geometry}, where a lens of mass $M_{\rm l}$ at redshift $z_{\rm l}$ deflects the signal from a source at redshift $z_{\rm s}$. The actual path followed by the signal\footnote{We recall that we are here working in the geometric optics limit.}, which is smoothly curved in the space-time surrounding the lens, can be - as a first approximation - replaced by two straight rays with a kink near the deflector. The difference between the angular position of the image and the angular position of the source is called deflection angle, and we denote it as $\hat{\alpha}$.\\
\indent The true position of the source is related to its lensed image on the sky through the lens equation, which reads \citep{Schneider:92aa} 
\begin{align}
    \pmb{\eta}=\frac{d_{\rm s}}{d_{\rm l}}\pmb{\xi}-d_{\rm ls}\pmb{\hat{\alpha}}(\pmb\xi).
    \label{eq:lens_1}
\end{align}
Taking into account the geometry illustrated in Figure \ref{fig:geometry}, the source and lens impact parameters can be written as
\begin{align}
    \pmb\eta=\pmb\gamma d_{\rm s},\,\,\,\,\, \pmb\xi=\pmb\theta d_{\rm l},
\end{align}
 and thus the Equation \eqref{eq:lens_1} becomes
 \begin{align}
 \pmb\gamma=\pmb\theta-\frac{d_{\rm ls}}{d_{\rm s}}\pmb{\hat{\alpha}}(\pmb{\theta}d_{\rm d})\equiv \pmb\theta-\pmb{\alpha}(\pmb\theta),
 \label{eq:lens_2}
\end{align}
where we have introduced the scaled deflection angle $\pmb{{\alpha}}=({d_{\rm ls}}/{d_{\rm s}})\pmb{\hat{\alpha}}$. The scaled deflection angle can be expressed in terms of the convergence $\ka$, as 
\begin{align}
    \pmb{\alpha}(\pmb{\theta})=\frac{1}{\pi}\int d^2\theta'\ka(\pmb{\theta'})\frac{\pmb{\theta}-\pmb{\theta'}}{|\pmb{\theta}-\pmb{\theta'}|},
\end{align}
where $\ka$ is defined as
\begin{align}
    \ka(\pmb\theta)=\frac{\Sigma(d_{\rm l}\pmb\theta)}{\Sigma_{\rm cr}}, \,\,\,\,\, \text{with}\,\,\,\Sigma_{\rm cr}=\frac{c^2d_{\rm s}}{4\pi Gd_{\rm l}d_{\rm ls}}\,.
\end{align}
Note that the surface mass density $\Sigma$ is obtained by integrating the mean mass density of the lens along the line of sight. 
We are in the strong lensing limit when $\Sigma > \Sigma_{\rm cr}$, i.e. when the mass distribution of the lens allows the production of multiple images of the source \citep[][]{Kochanek:06aa}.\\
\indent Let us now focus on the SIS model. The mass density of a SIS is given by \citep[][]{Narayan:96aa}
\be
\varrho(r) = \frac{\si^2_{\rm v}}{2\pi Gr^2} \,,
\ee
where $\si_{\rm v}$ denotes the velocity dispersion of the lens. Despite the singularity in the center  and the infinite total mass, this can be  considered as a rather realistic mass distribution for lensing by a galaxy (see, e.g., \citealt{Schneider:92aa}), with $\sigma_{\rm v}$ velocity dispersion within the galaxy. Integrating along the line of sight we obtain the surface density
\begin{align}
\Sigma(\theta)&=2\times \frac{\sigma_{\rm v}^2}{2\pi G}\int_{0}^{+\infty}\frac{1}{\xi^{2}+z^2} \nonumber\\
           &= \frac{\si_v^2}{2G}\frac{1}{\xi} \nonumber \\
           &= \frac{\si_v^2}{2G}\frac{1}{d_{\rm l}\th} \,.
\end{align}
Thus, for a SIS the convergence reads
\begin{align}
\ka(\th) = \frac{\Si(\th)}{\Si_{\rm c}} = 2\pi\frac{\si_{\rm v}^2}{c^2}\frac{d_{\rm ls}}{d_{\rm s}}\frac{1}{\th} ,
\end{align}
with a constant deflection angle
\be
\al(\th) =4\pi\frac{\si_{\rm v}^2}{c^2}\frac{d_{\rm ls}}{d_{\rm s}} =2\th\ka(\th)\equiv \al_0\,.
\ee
This is usually called \textit{Einstein angle} \citep[see, e.g.,][]{Schneider:92aa}, and in order to have multiple images of the source the following condition needs to be fulfilled $\gamma < \alpha_{0}$.\\
\indent If we rescale our variables by $\al_0$, we can define the rescaled image and source positions as $\bx=\pmb{\theta}/\al_0$ and $\by=\pmb{\gamma}/\al_0$, hence Equation \eqref{eq:lens_2} becomes
\begin{align}
    \by =\bx -\frac{\bx}{|\bx|}.
\end{align}
We can distinguish three cases: i) for $\by=0$ the solution is the Einstein ring $|\bx|=1$, ii) for $y=|\by|<1$ one solution is $x_1=|\bx_1|=1+y$ (on the same side of the line of sight as the source), the other one is  $x_2=|\bx_2|=1-y$ (on the opposite side), iii) for $y>1$ the second solution no longer exists.\\
\indent The Jacobian of the lens map is
\begin{align}
&A_{ij} = \delta_{ij}\left(1- \frac{1}{|\bx|}\right) +\frac{x_ix_j}{|\bx|^3}\,,\quad \nn\\
&{\rm det}A= 1- \frac{1}{|\bx|}\,, \qquad \mu = \frac{1}{|{\rm det}A|} = \frac{|\bx|}{|1-|\bx||} \,,
\end{align}
where we have formally introduced the magnification $\mu$. Expressing the total magnification of a point source at position $y$ in terms of $y$ we find
\be\label{e:muy}
\mu(y) = \left\{\begin{array}{ll} \mu(\bx_1) + \mu(\bx_2) =\frac{y+1}{y} +\frac{1-y}{y} =\mu_1+\mu_2=\frac{2}{y} \,, & y\leq1\,,\\
 \frac{y+1}{y} =1+\frac{1}{y} \,, & y\geq 1\,. \end{array} \right.
\ee
We observe that the magnification is always positive. This is a consequence of the fact that SIS is an overdensity, hence it cannot de-magnify the signal of a background source.

\subsection{Cross-section and optical depth}

The impact parameter of the source (in the source plane) is given by $|{\bf{\eta}}|=\eta=\gamma d_s=y\al_0d_s$. A source with impact parameter smaller or equal to $\eta$ is amplified by at least a factor $\mu(y)$. Hence, considering a SIS with velocity dispersion $\si_v$, the cross section for amplification $\geq\mu_1$ of the stronger image is
\be\label{e:cross}
\varsigma(\mu_1,z_{\ell},z_s,\si_v)= \pi \eta^2 =\pi(y\al_0d_s)^2 = \frac{\pi(4\pi)^2\si_v^4d_{\rm ls}^2}{c^4(\mu_1-1)^2}\,.
\ee
Note that this cross section gives the area, centered along the line of sight of the lens,  within which a source at $z_s$ must lie so that it is amplified by a factor $\mu_1$ or larger by the lens at $z_l$. 
The expression (\ref{e:cross}) remains valid also for $y\geq1$, where we have only one image with magnification $\mu_1$ which tends to $1$ when $y\ra\infty$.

In our study of strong lensing of gravitational waves  we consider only one image and not the sum of both, since we expect to see a short burst of GWs which comes only from one image. The second image is delayed in time, with typical time delay of the order of a few months (see e.g. ~\citealt{Oguri:2018muv}), much longer than the GW burst. Since we are interested in magnification, we shall compute the cross section for the stronger image. This point has been raised in \citet{Cusin:21aa} but it was neglected in the previous literature: usually in Equation\,\eqref{e:cross} $y^{-1}=\mu_1-1$ is replaced by $y^{-1}=\mu/2=(\mu_1+\mu_2)/2$ which is the correct expression for a static situation where both images are seen together. For strong amplification, $\mu_1\sim\mu_2\gg 1$ this difference reduces the cross section by a factor $4$.

To compute the corresponding optical depth, denoted $\tau(\mu,z_s)$, we  need to know the physical density $n(\si_v,z_{\rm l})$ of lenses (galaxies) with a given velocity dispersion $\si_{\rm v}$ at redshift $z_{\rm l}$. We define the density function $n(\si_{\rm v},z_{\rm l})$ such that its integral $\int n(\si_{\rm v},z_{\rm l})d\si_{\rm v}$ simply gives the total density of lenses at redshift $z_{\rm l}$. 

The optical depth for lensing with magnification $\geq\mu$ (of the most strongly magnified image in the case of two images)  for a source at redshift $z_{\rm s}$ is
\be\label{e:depth1}
\tau(\mu,z_{\rm s}) =\int_0^{z_{\rm s}}dz_{\rm l}\frac{dr}{dz_{\rm l}}\left(\frac{d_{\rm l}}{d_{\rm s}}\right)^2\int \varsigma(\mu,z_{\rm l},z_s,\si_{\rm v})n(\si_{\rm v},z_{\rm l})d\si_{\rm v}\,,
\ee
where $dr$ is the physical length element at redshift $z_{\rm l}$, while $n(\si_{\rm v},z_{\rm l})$ is the physical number density of lenses, which is related to the comoving number density by $n(\si_{\rm v},z_{\rm l})=(1+z)^{3}n^{\rm com}(\si_{\rm v},z_{\rm l})$.
Note also that we have rescaled the cross section to the lens redshift, $\varsigma \ra \varsigma(d_{\rm l}/d_{\rm s})^2$ since in eq.\,(\ref{e:depth1}) we multiply by the lens density at $z_{\rm l}$. 
Inserting this and \eqref{e:cross} for the cross section in Equation \,(\ref{e:depth1}) 
we obtain
\begin{align}\label{e:depth2}
&\tau(\mu,z_s) =\nn\\
&=\frac{\pi(4\pi)^2}{(\mu-1)^2}\int_0^{z_s}\!\!\!dz\frac{\chi^2(z,z_s)\chi^2(z)}{\chi(z_s)^2(1+z)^3H(z)}\int_0^\infty d\si_v\si_v^4 n(\si_{\rm v},z_{\rm l}) \,,
\end{align}
where $\chi(z_1,z_2)$ denotes the comoving distance from redshift $z_1$ to $z_2$. Eq.~\eqref{e:depth2}
 is the optical depth for magnification \emph{larger than} $\mu$.

Often only the strong lensing case is considered and the magnification from the two images is added to give the total magnification. To do this one has to replace $1/(\mu-1)^2$ by $4/\mu^2$. As already mentioned above, here we cannot do this since in the case of strong magnification and double images we expect a considerable time delay, so that typically we observe only one image at one given time. Here we assume this to be the stronger image. In the strong magnification case, $\mu\gg2$, this difference is roughly a factor $4$, while in the limit $y\ra 1$ where $\mu\ra 2$ and the second image disappears, the two expressions agree.

\section{SNR calculation}
\label{app:snr}
To calculate the SNR $\rho$ of a TDE for a given interferometer, we need to express the signal in the detector frame of reference
\begin{align}
    \rho&=\frac{h^{\rm d}_{\rm gw}}{h_{\rm c}(f^{\rm d}_{\rm gw})}=\\\nonumber
    &=\beta\times 2 \times 10^{-22}\times \left( \frac{M^{\rm d}_{*}}{M_\odot}\right)^{1/3}\times \left(\frac{M^{\rm d}_ {\bullet}}{10^6\text{M}_{\odot}}\right)^{2/3}\\\nonumber
    & \times\left(\frac{D}{16\text{Mpc}}\right)^{-1}\times \frac{1}{h_{\rm c}(f_{\rm gw}/(1+z))}.
\end{align}
For sake of simplicity, in the above formula we have considered the MS star case and already applied the mass-radius scaling relation of Equation \eqref{eq:mass_radius_ms}. Considering how to convert frequency and mass from detector to source-frame reference,
\begin{align}
      M^{\rm d}&=M(1+z),\\
    f^{\rm d}_{\rm gw}&=f_{\rm gw}/(1+z),
\end{align}
and the relation between comoving and luminosity distance \citep{Hogg:99aa}
\begin{align}
     \chi=\frac{D}{1+z},
\end{align}
we can re-write $\rho$ as presented in Equation \eqref{eq:snr}.

\bsp	
\label{lastpage}
\end{document}